\documentclass[a4paper,12pt]{article}

\usepackage{indentfirst}
\usepackage{amsfonts,amsmath,amssymb,bm,a4wide,graphicx,cite,makeidx,multicol,color,nicefrac,color}

\begin{document}

\title{Spin Light of neutrino in polarized matter}

\author{A.Grigoriev\footnote{ax.grigoriev@gmail.com}, \ A.Ternov
   \\
   \small {\it Moscow Center for Advanced Studies,}
   \\
   \small {\it 141701 20, Kulakova Str., Moscow, Russia }
   }

\date{}
\maketitle

\begin{abstract}
The Spin Light of neutrino ($SL\nu$) is electromagnetic radiation produced by the neutrino magnetic moment when a neutrino moves under external conditions, such as fields or matter. This effect may be significant in the extremely dense matter of compact astrophysical objects, such as neutron stars (NS). If detected, this radiation could provide a valuable opportunity to study the properties of neutrinos and of the medium through which they move, because the properties of the radiation depend on both. Motivated by the possibility of nuclear matter spin polarization, in this paper, we study the new properties of $SL\nu$ that arise under the influence of net matter polarization. We demonstrate that polarization can either enhance or completely suppress the radiation. It also introduces a characteristic asymmetry into the total radiation from the compact object, which could be an observable feature depending on the matter polarization and the magnetic field inside the star, if the field is connected to the stellar matter polarization. This research may have implications for the physics of NS and magnetars, bringing us closer to the possibility of studying their internal structure.
\end{abstract}


\section{Introduction}\label{intro}
The neutrino is one of the key topics in modern research on fundamental interactions and cosmology. This elementary particle is associated with direct evidence for the existence of ``New Physics'' beyond the Standard Model (SM) of elementary-particle interactions: the phenomenon of neutrino oscillations, for the definitive proof of which the 2015 Nobel Prize in Physics was awarded. In space physics and astrophysics, the neutrino plays an important role in many highly relevant problems, both in the recent past, such as the model of nuclear reactions in the Sun, and at present, including the mechanism of ultra-high-energy cosmic-ray generation and the theory of supernova explosions \cite{Bilenky2018,AtharPPNP2022}.

Because the neutrino interacts extremely weakly with other particles, its fundamental parameters remain largely uncertain; however, oscillation experiments clearly indicate the existence of neutrino mass. In contrast to the SM, in which the neutrino is massless and does not participate in electromagnetic interactions, a nonzero mass provides grounds for assuming the presence of nontrivial electromagnetic properties \cite{FujShroPRL1980}. Thus, even in the simplest extension of the SM with neutrino mass (the so-called minimally extended Standard Model), the Dirac neutrino has a magnetic moment.
\begin{equation}
\mu=\frac{3eG_{\mathrm{F}}m}{8\sqrt{2}\pi^{2}}
\simeq3.2\times10^{-19}\mu_{\mathrm{B}}\,\left(  \frac{m}{1\,\mathrm{eV}}\right)  ,
\label{1-AMM-Classic}
\end{equation}
where $e$ is the absolute value of the electron charge, $m$ is the neutrino mass, and $\mu_{\mathrm{B}}=e/2m_{e}$ is the Bohr magneton. The neutrino electric charge is expected to be vanishingly small; it remains exceptionally small even within the framework of more specific theories beyond the SM \cite{GiunStud_2015}. Current experimental constraints on the neutrino magnetic moment are at the level of $\mu\leq(2.8{-}2.9)\times10^{-11}\mu_{\mathrm{B}}$ \cite{GEMMA:2012,Borexino-AMM:2017}, while more stringent constraints have been obtained from astrophysical data: $\mu\leq(1.1{-}2.6)\times10^{-12}\mu_{\mathrm{B}}$ \cite{Viaux-clustM5-PRL:2013,Arceo-Diaz-clust-omega:2015,Kuznetsov:2009zm}.

Although apparently very small, the neutrino magnetic moment can lead to important phenomenological consequences. A well-known example is neutrino spin oscillations and spin-flavour oscillations, which are associated with changes in neutrino helicity under external conditions, such as a magnetic field or a matter environment. This effect has previously been invoked in attempts to explain the solar-neutrino problem, and on its basis a satisfactory theory of Type II supernova explosions is still being sought.

Another phenomenon directly related to the magnetic moment is the $SL\nu$ in matter,
which has been studied in a series of papers \cite{LobStud_PLB564,StudTern_PLB2005,Lob_PLB619,Grig-Stud-Ternov-PLB:05,Gr-Lok-St-Ternov-PLB:12}. The effect consists of the emission of electromagnetic waves by the neutrino magnetic moment during transitions of neutrinos between their quantum states in a medium. Owing to the net interaction with background particles, the corresponding energy levels become spin-dependent, and the energy gap arising between them powers the process. As a result, during the transition the neutrino changes its helicity, which gives the effect its name. This radiation has been studied in detail for various types of astrophysical media \cite{Gr-Lok-St-Ternov-JHEP:17}. It has been shown that it can be sufficiently effective to be of potential interest under certain conditions associated with compact astrophysical objects, such as proto-neutron stars of supernovae, neutron stars, gamma-ray bursts, and third-family compact stars. This circumstance is connected with the fact that under the conditions of the indicated environments the density reaches its highest values, which, together with the assumption of ultra-high neutrino energy, makes it possible to "compensate" for the extremely small value of the neutrino magnetic moment. These settings are also interesting because, within them, the $SL\nu$ has specific polarization properties that can be used to distinguish this kind of radiation from other types.

The most promising objects for the strongest manifestation of the $SL\nu$ have been shown to be neutron stars (NSs) and gamma-ray bursts (GRBs) \cite{Gr-Lok-St-Ternov-ICHEP:18,Gr-Lok-St-Ternov-JPA:20}. For instance, in the most optimistic scenario, the emission time in a NS may attain the value $\tau_{SL\nu}\simeq 300$~s for a neutrino energy of $E_\nu\simeq 10$~PeV. The relevant astrophysical sites may include galaxy clusters, which are thought to produce neutrinos of ultra-high energies \cite{Berezinsky-clusters:97}. It should also be noted that an even higher $SL\nu$ emission rate can be achieved in hypothetical third-family compact stars, whose density could exceed the nuclear saturation density $n_0=1.6\times 10^{38}$~cm$^{-3}$ characteristic of NSs.

The studies mentioned above contained a straightforward calculation of the $SL\nu$ process in uniform matter, where the number density was its only characteristic. However, more complex matter configurations may affect the neutrino states and alter the results. One such possibility is the relativistic motion of matter occurring in GRBs. In this case, the main feature added to the $SL\nu$ is amplification due to the Lorentz increase in the number density, $n \rightarrow \gamma n$, where $\gamma=1/\sqrt{1-v^2}$ and $v$ is the matter velocity \cite{GrigStudTern2024}. Owing to the quadratic dependence of the process on $n$, its efficiency can be increased almost tenfold.

Another factor that can affect neutrino motion and radiation in dense astrophysical matter is its (spin) polarization. This matter characteristic can have different origins, being either a source or a consequence of the magnetic field present in the astrophysical environment. This circumstance is especially relevant to magnetars, where magnetic fields are thought to reach magnitudes as high as $10^{18}-10^{19}$~G \cite{VidanaPRC2015}. If matter generates the field, the polarization is thought to be due to a ferromagnetic phase transition of nuclear matter (see \cite{HashimotoPRD2014,EtoPRD2013,VidanaPRC2016} and references therein). Another scenario utilizes concepts of some kind of dynamo mechanism and magnetic-flux conservation \cite{TurollaRPP2015} (and references therein) to obtain large magnetic fields, which then polarize the matter. We also point to the possibility of spin-polarized nuclear matter without necessarily generating a substantial magnetic field. For instance, Skyrme forces have been shown to manifest such a property under certain conditions \cite{KutscheraPLB1994,Perez-GarciaPRC2008}.

Substantial polarization and possible phase transitions in the matter of dense astrophysical objects, particularly magnetars, are widely discussed in connection with the elaboration of the equation of state (EOS) for hadronic matter at high densities, up to $n\simeq10n_{0}$ \cite{Dutra-Scirme:2012,Mikheev-Lanskoy(EOS):2022}. This problem is one of the most pressing and important in modern nuclear physics, and it is not solely of theoretical importance. Knowledge of such an equation is required to describe the observed properties of neutron stars, supernova explosions, and the emission of gravitational waves during the merger of binary neutron stars.

Over the past decade, our knowledge of neutron stars has expanded significantly: new mass and radius measurements have been made (see \cite{Koehn(EOS):2025} and references therein). The recent observation of the GW170817 event, conducted within the multi-messenger approach, was the first observation of a binary neutron star merger accompanied by the emission of gravitational waves, together with the detection of a short gamma-ray burst (see review \cite{Poggiani-GW170817:2025}). Analysis of the obtained results makes it possible to establish important constraints on the parameters of the EOS and allows for the emergence of spin-polarized nuclear matter in the inner regions of neutron stars, with a possible phase transition from the unpolarized to the spin-polarized state \cite{Tews-Schwenk-Ph-Trans:2020}. Moreover, a realistic scenario based on data from GW170817 suggests that up to 60\% of baryons have their spins polarized during a NS merger \cite{Tan-Khoa-60:2020}.

The polarization of matter modifies its net interaction with neutrinos and leads to changes in neutrino states. In this article, we describe these states and the new $SL\nu$ radiation properties that appear in this case. We will not discuss in detail how the field and matter polarization are distributed in compact objects; rather, we aim to assess the general features acquired by $SL\nu$ due to matter polarization. We would like to stress that since the magnetic moment for Majorana neutrinos is zero, the neutrino under study has to be of the Dirac type. Accordingly, below we consider the problem of Dirac neutrino motion and $SL\nu$ radiation in non-moving matter with uniform density and polarization. In addition, we do not address here how the matter acquires the polarization and instead treat it as an independent fixed quantity.

\section{Neutrino quantum states in polarized matter}\label{sec_2}
It is convenient to calculate the $SL\nu$ using the method of exact solutions \cite{StudTern_PLB2005,Stud:2008}, in which the initial and final neutrino states in matter are described by the modified Dirac equation \cite{StudTern_PLB2005,Lob_PLB619}. The transition between these states, accompanied by photon emission, is represented by the usual Feynman diagram with the standard magnetic-dipole electromagnetic vertex. The modified Dirac equation for a neutrino accounts for the net coherent interaction of the neutrino with matter particles and has the general form:
\begin{equation}\label{Dirac_eq}
  \left\{ i\gamma_{\mu}\partial^{\mu}- \frac{1}{2}\gamma_{\mu}(1+\gamma^5)f^{\mu} - m \right\}\Psi(x)=0,
\end{equation}
where, for $\gamma^5$, we follow the representation introduced in the textbook \cite{BerLifPit_1982}.
The 4-vector $f^{\mu}$ describes the neutrino interaction with matter and generally represents a linear combination of the 4-vectors $j^{\mu}_f$ of the matter current and the 4-vectors $\lambda^{\mu}_f$ of the matter polarization for each matter component $f$ (for ordinary matter, $f=e,\ p, \ n$). The coefficients in this combination are determined by the specific type of neutrino interaction with matter particles. The current and polarization vectors are given, respectively, by \cite{StudTern_PLB2005}
\begin{equation}
j_{f}^\mu=(n_f,n_f{\bf v}_f), \label{j}
\end{equation}
and
\begin{equation} \label{lambda}
\lambda_f^{\mu} =\Bigg(n_f ({\bm \zeta}_f {\bf v}_f ), n_f {\bm \zeta}_f \sqrt{1-v_f^2}+ \frac{{n_f {\bf
v}_f ({\bm \zeta}_f {\bf v}_f )}}{{1+\sqrt{1- v_f^2}}}\Bigg),
\end{equation}
where $n_f$ is the number density, ${\bf v}_f$ is the net velocity of component $f$, and ${\bm \zeta}_f$ ($0\leqslant |{\bm \zeta}_f | \leqslant 1$) is its average polarization in the rest frame. Here, ${\bm \zeta}_f {\bf v}_f $ denotes the scalar product, with analogous notation used for other vectors below. In the considered case of non-moving matter, Eqs. (\ref{j}), (\ref{lambda}) reduce to $j_{f}^\mu=(n_f,{\bf 0})$, $\lambda_f^{\mu} =(0,n_f {\bm \zeta}_f)$.

When choosing parameter values, we focus on magnetar matter. As neutron stars, magnetars consist mainly of neutrons, whose number density conventionally reaches values of up to $n_n \sim 10^{38}-10^{39}$~cm$^{-3}$ \cite{Weber-Book:99,Schmitt-DenseMat-Book:2010}. Some investigations suggest even higher densities in the NS interior, up to $10^{41}$~cm$^{-3}$ \cite{Belvedere:2012}. The fractions of protons and electrons are much lower and are typically considered to be at the level of $0.05n_n-0.1n_n$. There are indications that neutrons can have very high spin polarization, which can even be complete
\cite{IsaevPRC2009}. However, there is still no general agreement on this issue, and different authors have obtained conflicting results. Recent studies of nuclear matter in a strong magnetic field tend to conclude that polarization decreases as density increases \cite{VidanaPRC2014}. On the other hand, there is currently no consensus on when exactly the polarization becomes substantially low because of discrepancies among nuclear-force models and their incompleteness. Taking these facts into account, in what follows, we assume that the charged matter component is sufficiently low that its contribution to the matter current and polarization can be omitted and that the matter polarization $|{\bm \zeta}|$ takes values in the full range from 0 to 1. Finally, if we consider an electron Dirac neutrino with interactions within the minimally extended Standard Model, then, in the above approximation, the vector $f^{\mu}$ is found to be:
\begin{equation}\label{f_n}
  f^{\mu}=\frac{G_F}{2\sqrt{2}}\left( -n_n, n_n {\bm \zeta} \right).
\end{equation}
This expression will be used in our study. If the polarization of matter is not due solely to neutrons, the expression for $f^{\mu}$ can also be written in the form of Eq.~(\ref{f_n}) by redefining $n_n$ and ${\bm \zeta}$.

We first analyze the spectrum of possible neutrino states in polarized matter that follows from equation (\ref{Dirac_eq}). In the momentum representation and using (\ref{f_n}), the equation reads:
\begin{equation}\label{Dirac_final}
  \left\{ \gamma_{\mu}p^{\mu}+ \tilde{n}(\gamma^{0}+{\bm \zeta} {\bm \gamma})(1+\gamma^5) - m \right\}\Psi=0,
\end{equation}
where we have introduced the notation $\tilde{n}=G_F n_n/2\sqrt{2}$. To distinguish the states, all quantum numbers must be specified. For this purpose, we write the corresponding Hamiltonian (${\bm \alpha}=\gamma^0{\bm \gamma}$):
\begin{equation}\label{H}
  \mathrm{H}=({\bm \alpha}{\mathbf{p}})-{\tilde{n}}({\bm \alpha}{\bm \zeta})+\tilde{n}({\bm \Sigma}{\bm \zeta}) - \gamma^5\tilde{n} - \tilde{n} + \gamma^0m .
\end{equation}
First, we note that it commutes with the momentum, and therefore its value is a conserved quantum number. Determining the spin integral of motion is a nontrivial task, which was solved for the analogous problem in \cite{GrigStudTernovEPJC2022}. Following the approach developed there, we adopt the four-vector spin-polarization operator \cite{Sokolov-Ternov-Rel-El}, with the four-vector momentum replaced by the ``extended'' momentum $\tilde{p}^{\mu} \equiv p^{\mu}-f^{\mu}$:
\begin{equation}\label{T_mu}
  \widetilde{\mathrm{T}}^{\mu}=\gamma^5(\gamma^{\mu} - \tilde{p}^{\mu}/m).
\end{equation}
We then construct the scalar product $(\widetilde{\mathrm{T}} f)=\widetilde{\mathrm{T}}^{\mu}f_{\mu}$ and, using an appropriate coefficient, introduce the operator
\begin{equation}\label{S}
  \mathrm{S}=-(\nicefrac{m}{\tilde{n}}) \widetilde{\mathrm{T}}^{\mu}f_{\mu} = \gamma^5 \left[m\gamma^0 + m ({\bm \zeta}{\bm \gamma}) -  \widetilde{\mathrm{H}} - (\tilde{{\mathbf{p}}}{\bm \zeta})
  \right],
\end{equation}
where, according to the definition of the ``extended'' momentum, $\widetilde{\mathrm{H}}= \mathrm{H} + \tilde{n}$ and $\tilde{{\mathbf{p}}} = \mathbf{p} - \tilde{n}{\bm \zeta}$.

The resulting operator $\mathrm{S}$ commutes with the Hamiltonian and, together with the momentum operator, defines stationary states. Its eigenvalues are the observable values of the projection of the spin 4-vector onto $f^{\mu}$ in these states. To determine them explicitly, we write the operator in block form
\begin{equation}\label{S_block}
  \mathrm{S}=\begin{pmatrix}
      - m ({\bm \sigma}{\bm \zeta}) & - m + P \\
      m + P & m ({\bm \sigma}{\bm \zeta}) \\
    \end{pmatrix},
\end{equation}
where $\sigma_i$ denote the Pauli matrices, and the notation $P=(\tilde{p} f)/\tilde{n}$ is used. From (\ref{S_block}), the eigenvalues for solutions of Eq.~(\ref{Dirac_final}) are readily obtained and can be represented as $s\Lambda$, where the value of the spin observable is
\begin{equation}\label{Lambda}
   \Lambda = \sqrt{P^2-m^2(1-{\bm \zeta}^2)},
\end{equation}
and $s=\pm 1$ is the spin quantum number.

Taking into account the spin operator (\ref{S}), we obtain from the Hamiltonian (\ref{H}) the dispersion relation for a neutrino in polarized matter as follows:
\begin{equation}\label{Dispersion}
  p^2-m^2=2\tilde{n}(P-s\Lambda),
\end{equation}
where $p^2=E^2-{\mathbf{p}}^2$. The expanded form of this equation,
\begin{equation}\label{Dispersion_expanded}
  (p^2-m^2)^2-4(pf)(p^2-m^2)+4p^2\tilde{n}^2(1-{\bm \zeta}^2)=0,
\end{equation}
is a fourth-order algebraic equation that has no trivial solutions in the general case. The exact solution of Eq.~(\ref{Dispersion_expanded}) can be obtained for longitudinal matter polarization, ${\bm \zeta} {\parallel} \mathbf{p}$:
\begin{equation}\label{E_v||p}
  E_{||}=\varepsilon
            \sqrt{
                 (\mathrm{p}\mp\tilde{n}\zeta + s\tilde{n})^2+m^2
                 }
            \pm s\tilde{n}\zeta - \tilde{n},
\end{equation}
where the upper (lower) sign corresponds to the neutrino momentum directed along (opposite to) the matter polarization. The quantity $\varepsilon = \pm 1$ is the quantum number of the ``energy sign''. It divides the solutions into two branches, which, in the limit of vanishing matter density, reduce to the positive- and negative-frequency solutions of the Dirac equation in vacuum. For arbitrary orientations of the neutrino momentum and matter polarization, only approximate solutions can be given. For instance, when the neutrino mass is the smallest parameter, the dispersion relations for a negative-helicity neutrino $\nu_{s=-1}$ ($\approx \nu_L$) and a positive-helicity antineutrino ${\bar \nu}_{s=+1}$ ($\approx{\bar \nu}_R$) can be expressed as follows:
\begin{equation}\label{E_small_m}
  E=
     \varepsilon\sqrt{
          (\mathbf{p} - 2\tilde{n}{\bm \zeta})^2+m^2
          }
          - 2\tilde{n}.
\end{equation}
The other two solutions (for $\nu_{s=+1} \approx \nu_R$ and ${\bar \nu}_{s=-1} \approx{\bar \nu}_L$) in the same approximation have the trivial free-particle dispersion $ E=\varepsilon \mathrm{p}$.

Thus, neutrino states in polarized matter are specified by the momentum and the quantum numbers $\varepsilon$ and $s$. An exact solution of Eq.~(\ref{Dirac_final}) for the wave functions $\Psi_{\mathbf{p},\varepsilon,s}$ is impeded by the lack of a simple solution to the general dispersion relation (\ref{Dispersion_expanded}). The approximate solution needed to calculate the matrix element of the $SL\nu$ can be found by taking the following into account.
The current bound on the electron neutrino mass is $m_{\nu_e}\lesssim 0.45$~eV \cite{KATRIN2024}. As shown in previous studies, the efficiency of the $SL\nu$ increases with matter density and, of course, with neutrino momentum. Therefore, we are interested in the highest possible density values, $n_n\gtrsim 10^{38}$~cm$^{-3}$, which correspond to $G_F n_n \gtrsim 10$~eV on the energy scale. Comparing this value with the neutrino mass bound and taking into account that the neutrino is ultra-relativistic, we conclude that the approximation used in Eq.~(\ref{E_small_m}) is well justified in our problem. In this paper, we focus on effects at zeroth order in the neutrino mass and this quantity will be neglected in the calculations below.

One can also note that the zeroth component in (\ref{f_n}) is negative. This reflects the well-known fact that stable states in neutron matter are formed by {\it antineutrinos}, which therefore emit the spin light in our study. However, for convenience, we refer to them as {\it neutrinos} when no misunderstanding can arise. To rewrite the formalism above in terms of the antineutrino field, one simply has to change the sign of $n_n$  (and, accordingly, of $\tilde{n}$). In particular, the energy spectrum of neutrino species in the massless limit can be written as:
\begin{align}
  & E =\varepsilon |{\bf p}+2\tilde{n}{\bm \zeta}|+2\tilde{n}, \text{  \ for } \nu_L \text{\ and \ } {\bar \nu}_R \label{Disperion_active} \\
  & E =\varepsilon \mathrm{p}, \text{  \ for } \nu_R \text{\ and \ } {\bar \nu}_L. \label{Disperion_sterile}
\end{align}
The first relation applies to active neutrinos, whereas the second applies to sterile neutrinos. In the present problem, the initial state is represented by ${\bar \nu}_R$, with energy $E_i=|{\bf p}+2\tilde{n}{\bm \zeta}|+2\tilde{n}$, and the final state ${\bar \nu}_L$ is sterile, with $E_f=\mathrm{p}$.

In unpolarized matter, the expression for the energy of an active neutrino reduces, in the massless limit, to the well-known formula $E=\mathrm{p}+2\tilde{n}$. The expression for $E_i$ shows that matter polarization can considerably enhance the matter potential. In highly polarized matter with longitudinal neutrino motion, ($|{\bm \zeta}|\rightarrow 1$, ${\bm \zeta} || {\bf p}$), it is almost twice as large as in an unpolarized medium.

The solutions of equation (\ref{Dirac_final}) have the standard plane-wave form $\Psi =u({\bf p})e^{-ipx}$. To simplify the calculation of the $SL\nu$ amplitude below, we choose the coordinate system such that the initial neutrino moves along the {\it z} axis and the polarization lies in the {\it xz} plane. In this particular case, the normalized spinor corresponding to the initial neutrino with energy $E_i$ is obtained as follows (the standard representation of the $\gamma$-matrices is used throughout the paper):
\begin{equation}\label{Psi-i}
  u_i({\bf p}) = \frac{1}{\sqrt{2}\,L^{3/2} \sqrt{\mathrm{p}^2 + \tilde{n}^2\zeta^2\sin^2\delta}}\left(\begin{array}{c}
 -\tilde{n}\zeta\sin\delta
\\
 \mathrm{p}
\\
  \tilde{n}\zeta\sin\delta
\\
 -\mathrm{p}
\end{array}\right),
\end{equation}
where $\delta$ is the angle between the initial neutrino momentum ${\bf p}$ and the matter polarization ${\bm \zeta}$, and $L$ is the normalization length. The spinor for the final sterile neutrino with energy $E_f$ propagating in an arbitrary direction is given by:
\begin{equation}\label{Psi-f}
  u_f({\bf p}) = \frac{1}{2\,L^{3/2}}\left(\begin{array}{c}
 \sqrt{1+\frac{\mathrm{p}_z}{\mathrm{p}}}
\\
 \sqrt{1-\frac{\mathrm{p}_z}{\mathrm{p}}}e^{i\varphi}
\\
  \sqrt{1+\frac{\mathrm{p}_z}{\mathrm{p}}}
\\
 \sqrt{1-\frac{\mathrm{p}_z}{\mathrm{p}}}e^{i\varphi}
\end{array}\right),
\end{equation}
where $\varphi$ is defined by $\tan\varphi=p_2/p_1$. Since the final neutrino is sterile, the corresponding solution is the free-particle spinor in the massless limit.

\section{Emission of the $SL\nu$ photon}

A description of the radiation process should begin by establishing the conditions under which it can occur. The fundamental constraint to consider is the energy–momentum conservation law:
\begin{equation}\label{Kinemat}
  E_i=E_f + \omega, \ \  {\bf p}^{\prime} = {\bf p} + {\bf k},
\end{equation}
where ${\bf k}$ is the momentum of the radiated photon. Owing to the presence of electrons, the photon acquires the properties of a massive plasmon, with energy $\omega=\sqrt{{\bf k}^2+m^2_{\gamma}}$. Since the electrons in a NS form a relativistic degenerate Fermi gas, the plasmon mass is given by \cite{Gr-Lok-St-Ternov-PLB:12}:
\begin{equation}\label{m_gamma}
  m_{\gamma}=(2\alpha)^{1/2}(3\sqrt{\pi}n_e)^{1/3}\simeq 8.9 \times\left(  \frac{n_{e}}{10^{37}\,\text{cm}^{\!-3}}\right)^{1/3}\text{MeV}.
\end{equation}
The plasmon mass scale is much larger than the possible values of the initial neutrino effective potential in matter, which is of the order of $\tilde{n}\lesssim 10^3$~eV. This clearly indicates the existence of a reaction threshold \cite{Gr-Lok-St-Ternov-PLB:12,Kuz-Mikh-anti:2007}.

Let us assume that the initial neutrino momentum is much larger than the matter potential. Then the initial energy can be approximately written as
\begin{equation}\label{E_i_approx}
  E_i \approx p+2\tilde{n}(1+\zeta \cos \delta),
\end{equation}
This differs from the case of unpolarized matter by the substitution $\tilde{n}\rightarrow \tilde{n}(1+\zeta \cos \delta)$. Thus, the threshold condition can be obtained straightforwardly by following the same procedure as in \cite{Gr-Lok-St-Ternov-PLB:12}, yielding:
\begin{equation}\label{Threshold}
   p_{th}=\frac{m^2_{\gamma}}{4\tilde{n}(1+\zeta \cos \delta)}<p.
\end{equation}
The polarization dependence makes the threshold condition more restrictive when the polarization is directed opposite to the neutrino momentum and less restrictive in the opposite case. In the latter case, for typical values ($n_e \sim 10^{37}~\text{cm}^{-3}$, $n_n \sim 10^{38}~\text{cm}^{-3}$), the threshold momentum is estimated by order of magnitude to be $p_{th} \sim 10~\text{TeV}$. This value is much greater than possible values of $\tilde{n}$, so the condition $\tilde{n} \ll p$ is satisfied. Such large values of $p\approx E_i$ are precisely why specific astrophysical settings with ultra-high-energy neutrinos are considered, as discussed in the introduction with reference to \cite{Gr-Lok-St-Ternov-JHEP:17}.

A solution to relations (\ref{Kinemat}), taking into account (\ref{E_i_approx}), can be readily found. However, for our purposes, it is sufficient to use its simple form when the condition $m^2_{\gamma}/4\tilde{n}p \ll 1$ is fulfilled (the far-above-threshold regime). In this case, one can neglect the dependence on $m_{\gamma}$ in the photon energy to obtain
\begin{equation}\label{omega}
  \omega = \frac{2p\tilde{n}(1+\zeta \cos \delta)}{p(1 - \cos \theta) + 2\tilde{n}(1+\zeta \cos \delta)},
\end{equation}
where $\theta$ denotes the angle between the propagation directions of the initial neutrino and the photon.

The coupling of an $SL\nu$ photon to a neutrino is described by the phenomenological Lagrangian for the interaction between the Dirac and electromagnetic fields via the anomalous magnetic moment:
\begin{equation}\label{L}
  {\cal L}= -\frac{1}{2}\mu~\overline{\nu}_{\alpha}\sigma_{\mu \nu}\nu_{\beta}F^{\mu \nu},
\end{equation}
where $\sigma_{\mu \nu}=i/2 (\gamma_{\mu}\gamma_{\nu}-\gamma_{\nu}\gamma_{\mu})$, $\gamma_{\mu}$ are the Dirac matrices, and $F^{\mu \nu}$ is the electromagnetic field tensor. The amplitude for the process, after integration over the space-time coordinates, then takes the form:
\begin{equation}\label{Sfi}
   S_{f i}=
  \mu~(2\pi)^4{\sqrt {\frac {2\pi}{\omega L^{3}}}}~\delta(E_f-E_i+\omega)\delta^3({\mathbf p}^{\prime}-{\mathbf p}+{\mathbf k})
  {\bar u}_{f}({\mathbf p}^{\prime})({\hat {\bm \Gamma}}{\mathbf e}^{*})
  u_{i}({\mathbf p}).
\end{equation}
Here, ${\mathbf e}$ is the photon polarization vector, and $\hat{\bm \Gamma}$ is the electromagnetic dipole vertex
\begin{equation}\label{Gamma}
  \hat {\bm \Gamma}=i\omega\big\{\big[{\bm \Sigma}
  {\bm \varkappa}\big]+i\gamma^{5}{\bm \Sigma}\big\},
\end{equation}
where ${\bm \varkappa}={\mathbf k}/{\omega}$ and ${\bm \Sigma}=\gamma^0 \gamma^5 {\bm \gamma}$.

The differential width is obtained from Eq.~(\ref{Sfi}) using the standard technique, which yields:
\begin{equation}\label{dGamma}
   d\Gamma=-\frac{{\mu}^2}{16\pi\omega}
      \delta(E_f+\omega -E_i)
      \delta^3({\mathbf p}^{\prime}+{\mathbf k}-{\mathbf p})
      |({\mathbf e}^*, {\mathbf j}_{fi})|^2
      d^3{\mathbf k}d^3{\mathbf p}^{\prime},
\end{equation}
where the neutrino current is ${\mathbf j}_{fi}({\mathbf k},~{\mathbf p}^{\prime})={\bar u}_f {\hat{\bm \Gamma}} u_i$. The total rate is obtained from this formula by integrating, using $k=\omega$, where $\omega$ is given by (\ref{omega}). However, the resulting expression is too cumbersome; therefore, we expand it in the small parameter $\tilde{n}/p\ll 1$ and retain the first two nonzero orders:
\begin{equation}\label{Gamma_Total}
  \Gamma=
  4\mu^{2}\tilde{n}^{2} (1+\zeta\cos\delta)\left( p(1+\zeta\cos\delta) + \tilde{n} (3 \zeta^2 \cos^2\delta + 4 \zeta\cos\delta -\zeta^{2}+2)\right).
\end{equation}

\begin{figure}[t]
 \centering
 \begin{tabular}{c}
      \begin{minipage}{0.5\hsize}
        \begin{center}
          \includegraphics[width=6cm,height=4cm]{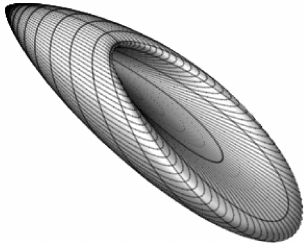}
        \end{center}
      \end{minipage}

      \begin{minipage}{0.5\hsize}
        \begin{center}
          \includegraphics[width=8cm]{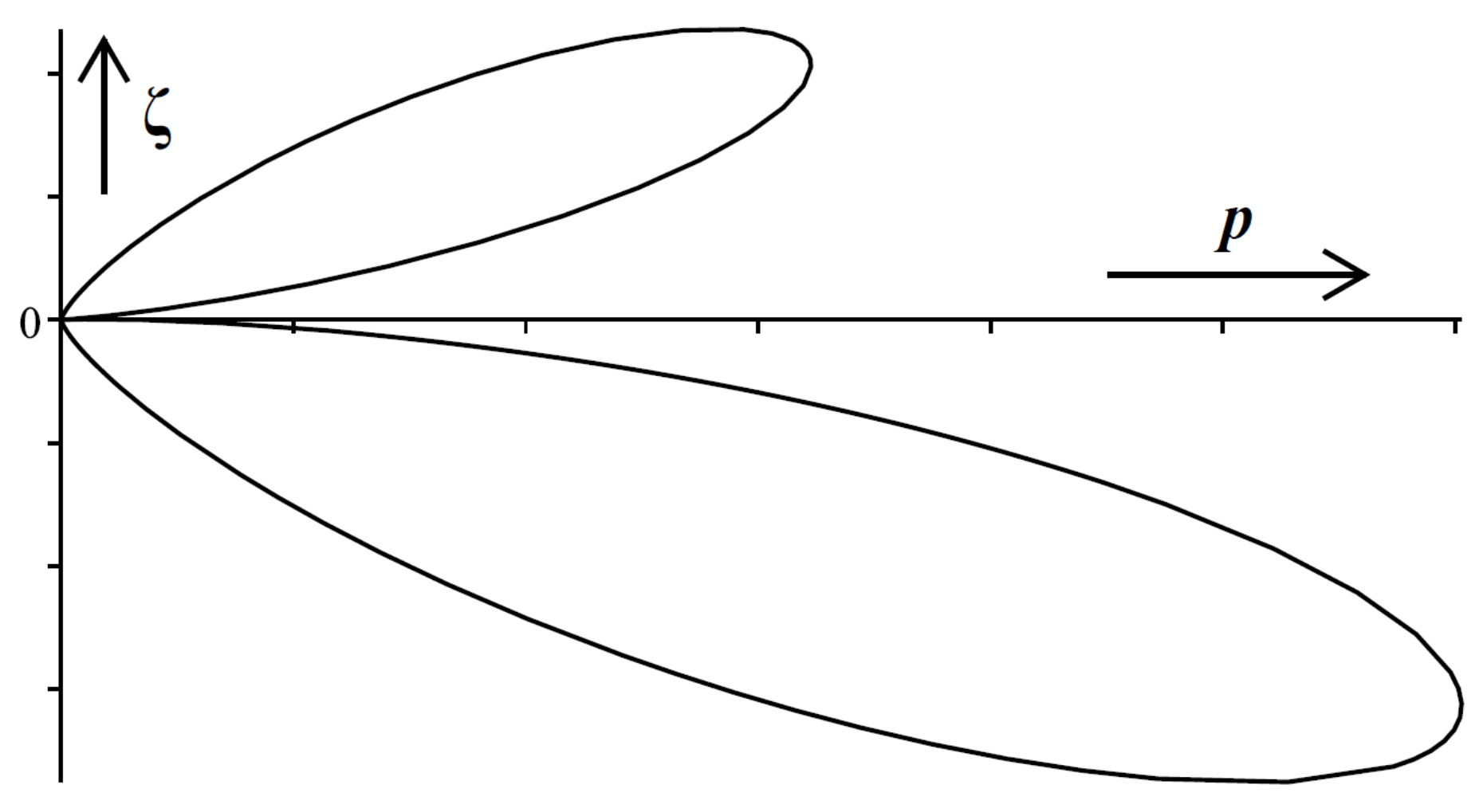}
        \end{center}
      \end{minipage}
 \end{tabular}
{\caption{Angular distribution of the radiation power (in arbitrary units) for the parameter set $\delta = \pi/2$, $\zeta = 1$, $\varphi = 0$, and $\tilde{n}/p = 0.01$: (a) the three-dimensional view; (b) the two-dimensional cross-section in the plane defined by the vectors ${\bm \zeta}$ and ${\bf p}$. \label{Angular distribution}}}
\end{figure}

The radiation power is calculated from the differential width as $dI=\omega d\Gamma$. Integrating  this relation and applying the same expansion yields the angular distribution of the radiation power:
\begin{equation}\label{dIdOmega}
  \frac{dI}{d\Omega}
  =\frac{16}{\pi}\tilde{n}^4 p^3\mu^{2}(1+\zeta\cos\delta)^{4}
       \frac{ (p (1-\cos\theta)-2\zeta\tilde{n}\cos\varphi \sin\theta \sin\delta) }
            {\left(p(1- \cos\theta)+2 \tilde{n} (1+ \zeta \cos\delta)\right)^4}.
\end{equation}
The dependence on the azimuthal angle $\varphi$ indicates the presence of radiation asymmetry caused by matter polarization. To illustrate this effect, we plot the angular distribution of the radiation in Fig.~(\ref{Angular distribution}) for the polarization direction fixed by $\delta = \pi/2$, for which the asymmetry is maximal. In addition, the asymmetry vanishes as $\tilde{n}/p \rightarrow 0$ and is most evident for the smallest possible $p$; therefore, for the plot, we choose $\tilde{n}/p = 0.01$.

The total radiation power obtained from (\ref{dIdOmega}) is:
\begin{equation}\label{I}
  I=\frac{4}{3} \mu^{2}\tilde{n}^{2} p \ (1+\zeta\cos\delta) \left(p(1+\zeta\cos\delta)+2\zeta^2 \tilde{n} \sin^{2}\delta\right).
\end{equation}
Substituting $\zeta=0$ into Eqs.~(\ref{Gamma_Total}) and (\ref{I}), we recover the results of our previous studies \cite{Grig-Stud-Ternov-PLB:05} for unpolarized matter in the same small-$m$ and small-$\tilde{n}$ limit as above ($m/p\ll \tilde{n} \ll p/m$):
\begin{equation}\label{Gamma_I_zeta_0}
  \Gamma=4 \mu^{2} \tilde{n}^{2} p, \ \ \  I=\frac{4}{3} \mu^{2}\tilde{n}^{2} p^2.
\end{equation}
The characteristic feature of the $SL\nu$ in polarized matter is the factor $(1+\zeta\cos\delta)$ in the derived formulas, which can either reduce or enhance the effect. In the most favorable case, when the matter is fully polarized along the initial neutrino propagation direction, this factor reaches the value $2$. In the considered limit $p\gg\tilde{n}$, this leads to a fourfold increase in the width and power. Conversely, if a neutrino moves against the matter polarization, this factor completely suppresses the radiation.

As pointed out in our previous studies \cite{Gr-Lok-St-Ternov-JHEP:17}, the polarization properties can be important for experimental studies of the $SL\nu$. In the case of vanishing neutrino mass, applying the analysis implemented in \cite{StudTern_PLB2005} and decomposing the photon polarization vector into the corresponding components, we obtain that the $SL\nu$ does not exhibit preferred linear polarization. At the same time, the radiation is circularly polarized: the right-circular component is absent (the radiation power is zero), and all emission is due to left-circularly polarized photons. This property is convenient for experimental identification of the $SL\nu$ radiation from astrophysical sources against the background of other radiation types.

\section{Conclusions and Discussion}
The main effect of matter polarization on $SL\nu$ is the appearance of the factor $(1+\zeta\cos\delta)$, which establishes the correlation between the radiation and the polarization direction. This feature can be discussed from two perspectives. For a single neutrino, the most efficient radiation occurs in a completely polarized along the neutrino momentum matter. In this case, the radiation time for an ultra-high-energy electron neutrino with $E \simeq p = 10$~PeV, $\mu \simeq2.9\times10^{-11}\mu_{B}$, and $n_n=10^{38}~\text{cm}^{-3}$ is estimated from (\ref{Gamma_Total}) to be
\begin{equation}
  \tau_{\mathrm{SL}\nu}=1/\Gamma \simeq 20~\text{s}.
  \label{5-LifTimeSLN}
\end{equation}
When the polarization vector tends toward the opposite direction, this factor approaches zero, and the radiation disappears. This behavior is connected with the second possible manifestation of the indicated feature: the dependence of the total neutrino spin radiation emitted by a compact source on the orientation of the matter polarization in the source, specifically on the angle between the directions of observation and matter polarization. Thus, we predict an asymmetry in the spin light radiation of neutrinos moving in a compact astrophysical source if its matter is polarized.

Along with the radiation itself, the process under study is accompanied by the transition of neutrinos into the sterile state $\bar{\nu}_L$, as indicated above. The rate of sterile-neutrino generation can be estimated once the density of the initial neutrino flux is specified. This process may be supplementary to conventional neutrino conversion into sterile states through oscillations, since both change the active--sterile composition of the neutrino flux. The evaluation of the phenomena discussed here for compact astrophysical sources requires numerical modeling based on plausible assumptions about the matter polarization configuration and the high-energy neutrino flux distribution within the source, and is planned for future work.

\section{Acknowledgements}
The authors thank M. V. Zverev for valuable discussions.

\end{document}